# Point transformer for protein structural heterogeneity analysis using CryoEM

Muyuan Chen [1*], Muchen Li [2,3], Renjie Liao [2,3,4]

[1] Division of CryoEM and Bioimaging, SSRL, SLAC National Accelerator Laboratory, Stanford University,
[2] Department of Electrical and Computer Engineering, The University of British Columbia
[3] Vector Institute
[4] Canada CIFAR AI Chair

* Correspondence to: muyuanc@stanford.edu

**Abstract**

*Structural dynamics of macromolecules is critical to their structural-function relationship. Cryogenic electron microscopy (CryoEM) provides snapshots of vitrified protein at different compositional and conformational states, and the structural heterogeneity of proteins can be characterized through computational analysis of the images. For protein systems with multiple degrees of freedom, it is still challenging to disentangle and interpret the different modes of dynamics. Here, by implementing Point Transformer, a self-attention network designed for point cloud analysis, we are able to improve the performance of heterogeneity analysis on CryoEM data, and characterize the dynamics of highly complex protein systems in a more human-interpretable way.*

**Introduction**

The structural dynamics of macromolecules is critical to their functionalities. To fulfill their roles in life activities, proteins are constantly undergoing dynamic structural changes inside cells to interact with their surrounding environment [1–3]. Better understanding of the conformational changes of proteins will lead to insights on their structure-function relationship and the mechanism of various biological processes.

Cryogenic electron microscopy (CryoEM) images proteins sample frozen in vitrified ice, and produces snapshots of many copies of the protein, i.e., particles, at their different states and in random orientations [4,5]. The orientations of the particles are determined computationally, and the 3D structures of the protein are reconstructed from those particles at as high as atomic resolution [6,7]. The structural heterogeneity of the sample has long been a major obstacle for high-resolution structure determination with CryoEM. However, the flexibility of the sample also provides an opportunity for understanding of the dynamics of the system, if the heterogeneity within the sample can be analyzed comprehensively using advanced computational methods [8–12].

To address the problem of protein structural heterogeneity, we have previously introduced a computational method that uses Gaussian mixture models (GMMs) for protein structure representation [13], and a pair of deep neural networks (DNNs), an encoder and a decoder, are used to embed the particle images to a low dimensional space corresponding to their conformations through unsupervised training. Compared to traditional voxel-based maps, the GMM provides a more sparse representation of protein structures, greatly reducing required computation resources and speeding up convergence. Previously we have shown that the GMM based approach can produce visually identical results on benchmark datasets much faster, compared to other voxel-based DNN methods. Additionally, it has been shown that the learned conformation changes can be used to retrieve high-resolution information of flexible domains, and reveal structural features previously hidden due to the structural heterogeneity [10,14].

Despite the advantages of GMMs in protein structure representation, it also presents difficulties in DNN designs. Compared to image based DNN models that can take advantage of convolutional layers, learning GMM representations from image data is less straightforward. While simple multilayer perceptron (MLP) based models have achieved good results in simulated and experimental datasets, the location information, as well as hierarchical nature of protein dynamics is largely ignored. In a flexible protein system, large scale movement of individual domains drives the motion of secondary structure elements (SSEs) inside, and the conformational changes at the SSE level leads to the movement of individual residues and sidechains. Utilization of the geometry information can potentially reduce overfitting and improve the accuracy of the heterogeneity analysis [14,15].

**Method**

To take into account the geometry of proteins in CryoEM heterogeneity analysis, here we implemented a self-attention based DNN design [16], Point Transformer (PT), for solving the structural dynamics of protein complexes [17]. Transformer-based networks have revolutionized natural language processing over the past few years, and have shown advantages in image analysis tasks. PT applies the Transformer architecture to 3D point clouds, and has achieved improved performance in tasks including semantic segmentation and object classification. The point cloud targeted design makes it a natural fit for the GMM representation of proteins, and the advanced features PT provides, including self-attention and position encoding, can potentially incorporate the prior of protein geometry into the analysis of noisy particles.

To implement PT for CryoEM heterogeneity analysis, we kept the same training method and loss function, as well as the encoder architecture described in prior publications, and only replaced the MLP-based decoder with a PT-based one. In our experiments of various encoder-decoder based DNNs for CryoEM heterogeneity analysis, the encoder generally performs a less significant role, and can even be omitted in some cases and still achieve comparable results [8], while changing decoder architecture has a stronger impact on the performance of the model.

In the new decoder design, we use 3 layers of upsampling PT, with 64, 256, 1024 points in the layers. The positions of the points are generated through k-means clustering from the full GMM. Since the output of the decoder is the relative movement or amplitude change of each Gaussian in the GMM, the point cloud geometry remains relatively consistent for all particles of each dataset. As such, we pre-compute the points for the intermediate layers, as well as the neighbors for the self attention and up-sampling, to reduce the computational resource cost during training.

The architecture of the DNN is shown in Figure 1. Starting from the low dimensional latent space input from the encoder, we first convert the input to (B, 64, 256) tensor, where B is the batch size, through a two-layer MLP, and assign the values to the 64 points in the first PT layer. The output from the final PT layer is upsampled to the final GMM representation through the same transition up step in the PT, and then passed to another dense layer to reshape the output to the required channels of the GMM (3D coordinates, amplitude, and sigma). In addition to the standard PT architecture, we also introduced additional residual layers [18], which directly upsample the output of the first two PT layers to the points in the final GMM. These residual outputs are appended to the upsampled output of the last PT layer and fed into the final dense layer to generate the GMM. In our tests, the residual layers improve the speed of convergence and the accuracy of the model, particularly in examples of complex dynamics with multiple degrees of freedom.

**Results**

To test the performance of the PT based approach, we start from applying the method on simulated CryoEM datasets. Given the good performance of existing MLP based methods on simulated data, here we designed a complex system to show the differences (Figure 2A). The data is based on intermediate states of 50S ribosome assembly (EMPIAR-10076) [19], in which we segment the mature 50S ribosome into 9 pieces of variable sizes. Other than the largest central segment, each of the 8 pieces independently has a 50% chance to be absent in each particle. In sum, there are 256 different compositional states among the 12,800 particles, presenting a challenge for most heterogeneity analysis methods. The orientations of particles are evenly distributed, and are provided to the heterogeneity analysis method. Realistic noise and contrast transfer functions (CTF) are also simulated in the particles [20].

Because the GMM-based heterogeneity analysis outputs with amplitude for each Gaussian function, it is easy to directly compare the decoder output with the ground truth used for simulation. We match each Gaussian function to the closest atoms in the atomic model of ribosome used for simulation, and calculate the accuracy by comparing the Gaussian amplitude with the presence/absense of the corresponding atoms in that particle.

Comparing MLP and PT based analysis on the simulated ribosome dataset, PT produced more separable classes based on the UMAP visualization of the latent space, cleaner decoder output, as well as higher accuracy (96.1% vs 93.2%) than the MLP based method. Notably, PT based decoder generated GMMs with more localized, and more biologically reasonable, regions of compositional changes, and avoided the dust-like individual amplitude spikes from the MLP based decoder (Figure 2B). The improvement is likely a result of the position encoding and self-attention around local neighborhoods provided by the PT.

While the PT based models provide high accuracy decoder outputs, for complex protein systems with hundreds of conformations, it is still difficult to decipher the latent space and present all possible classes in a human interpretable way. Even with a perfect clustering algorithm that can divide the latent space into hundreds of classes, each individual class would have too few particles to generate good 3D reconstruction. While it is possible to focus the heterogeneity analysis on smaller regions of the molecule and limit the degrees of freedom, deciding which regions to focus on can still be challenging, and the decision may introduce additional bias.

To make the heterogeneity analysis more interpretable, we use a similar protocol to *Sheng et al. 2023* [21], but without the iterative subclassification process. The high accuracy of the PT decoder makes it possible to directly apply UMAP to the GMM outputs without the massive classification used in the previous work [21]. Here we take the decoder output from all particles, and embed all Gaussian functions of the GMM to a 2D space using UMAP [22]. In the resulting space, each point represents one Gaussian function, and clustering of the points represents groups of Gaussian functions that change their amplitude in a correlated way across many particles (Figure 3A-B). Here we generated clusters from the latent space using DBScan, and plotted the corresponding Gaussian functions for each class in the GMM. The resulting clusters recover the independent regions of the 50S ribosome that was used for the simulation. With this information, it is then possible to focus on any one of the regions, and embed all particles to a 2D space using only GMM outputs of the corresponding regions. Further clustering on the new latent space can produce classes of particles with only structural heterogeneity on the target region, which can be reconstructed to generate 3D maps highlighting the structural differences (Figure 3C).

Comparing the GMM latent space, the MLP based approach recovered 6 regions, whereas the PT based method can separate 8 of them correctly without any manual intervention. Therefore, using the PT-based method, it is possible to recover at least 128 classes from the simulated dataset.

Next, we evaluate the performance of the PT based method on simulated datasets with continuous domain movement. Here we simulate the CryoEM dataset based on an ABC transporter from EMPIAR-10374 [23]. Similar to the previous example, we segment the structure to four regions, and simulate independent continuous tilting motion for three of the four pieces (Figure 4A). To ensure a dataset with true continuous motion, each particle is independently simulated from one molecular model, which has the three local regions rotated by three random angles. For all particles, the region of movement and the axis of rotation are identical, and the random rotation angles are uniformly distributed. As such, there are three independent degrees of freedom within the dataset, and the average RMSD of per particle atomic movement is 2.23 Å.

Similarly, we apply both the MLP and PT based methods on the simulated dataset, and directly compare the per particle decoder GMM output with the coordinates in the ground truth atomic models to reflect the accuracy of the methods. On average, the PT based method achieved an RMSD of 0.67 Å, slightly better than the MLP based method (with a two-level hierarchical GMM), which has an RMSD of 0.70Å (Figure 4B). While the average improvement seems small, at the level of individual particles, the PT based model better captured the subtle movement from particle images (Figure 4C). Although the movement does not contribute much to the overall RMSD, they can still play important roles in biological processes in real datasets.

By embedding the Gaussian functions to 2D latent space using the decoder output, regions with independent movement modes become separable, and the segmentation of regions from the latent space matches well with the initial segmentation used in the simulation. To focus on one mode of movement, we selected one cluster and embedded the particles onto a latent space using only the information of Gaussian functions from that cluster. In the latent space, the particles form an 1D curve, depicting the simulated movement of that region (Figure 4D). By clustering particles on the different latent spaces, we can also reconstruct 3D maps that recover all three independent movement trajectories (Figure 4E). The divide-and-conquer approach makes it easier to analyze complex protein systems, where multiple parts are undergoing simultaneous conformational changes.

Finally, we apply the PT based model to real CryoEM dataset to show what new information we can learn using the method. Again, to construct a challenging case, we take the same 50S ribosome assembly dataset (EMPIAR-10076) [19], and select particles that land in between the clusters of the main conformations analyzed by existing methods (Figure 5A) [13]. The goal here is to see if additional intermediate states can be recovered from the "leftover" particles using the new method.

Comparing the results from the MLP and PT based analysis on the real dataset, the PT based approach produced more structured latent space and more meaningful decoder output with localized structural changes (Figure 5B). By mapping the GMM to a latent space based on decoder output, we can segment the ribosome structure into multiple pieces, some containing only a few turns of RNA helices. Subsequent classification and 3D reconstruction focusing on individual regions can then reveal the step-by-step addition of RNA segments not shown by previous analysis (Figure 5C-D).

**Discussion**

In sum, here we introduced a PT based architecture for CryoEM heterogeneity analysis. The model is built on top of our existing heterogeneity analysis method that uses GMMs for protein density representation and DNNs for learning conformations from particles. The self-attention and spatial encoding introduced by the PT layers greatly improved the performance of the analysis. By only swapping the decoder MLP layers with PT ones, the model produced a more separable latent space and more meaningful decoder outputs. Since the GMM based representation makes it possible to directly compare the decoder output to the molecular model used for the simulation, we also report ground truth accuracy of the trained models, which showed clear improvement of the PT based method. Application of the new method on real dataset also revealed subtle structural changes that were hidden from existing analyses.

To better interpret the results of DNN based heterogeneity analysis, especially for complex and dynamic protein systems with multiple degrees of freedom, we showed that the Gaussian functions in the GMM can be embedded to a latent space using the decoder output. By clustering on this latent space, we can segment the GMM into regions that have correlated dynamics. I.e. parts of a complex that get assembled as a single piece, or domains of a protein that move together as a rigid body. In contrast, separated clusters in the GMM latent space represent regions of the protein that have independent patterns of dynamics. By embedding all particles to a latent space using information on each individual cluster of the GMM, we can visualize the conformational change of the target region. The divide-and-conquer approach presents the dynamics of complex protein systems in an interpretable way to human researchers. Through this method, we can easily show which parts of the protein are moving, the movement patterns of the individual pieces, and how the movement of different domains are coordinated.

**Code availability**

The PT based heterogeneity analysis tool is distributed with the EMAN2 software package, and can be accessed using the "e2gmm_refine_jax.py" program with the "--pointtransformer" option.

**Acknowledgement**

This research is supported by NIH grant R01GM150905.

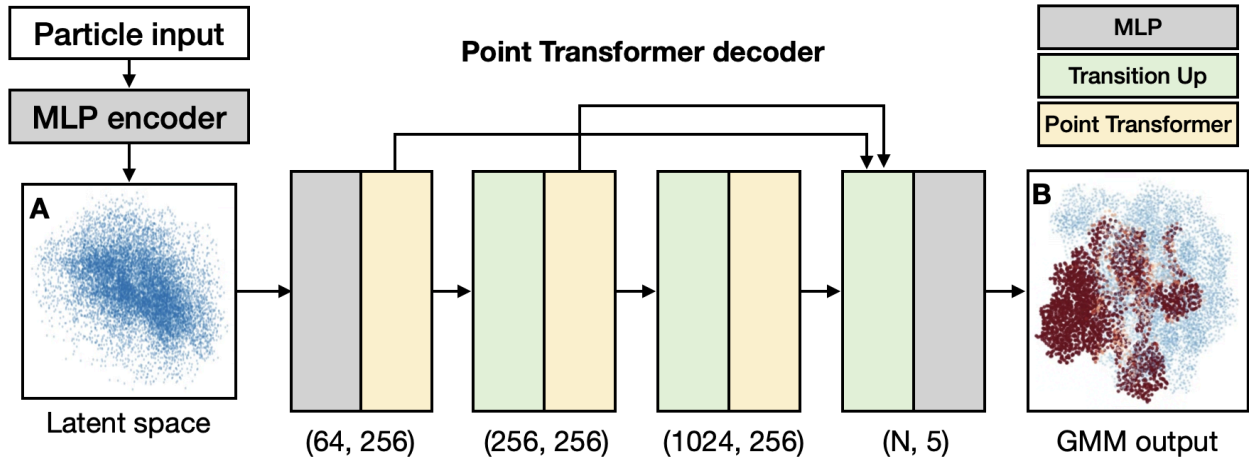

Figure 1. Architecture of the PT-based DNN for CryoEM structural heterogeneity analysis. (A) Latent space of the simulated ribosome dataset, showing the first two dimensions. Each point corresponds to one 2D particle input. (B) GMM representation of a ribosome particle generated by the decoder. Each point is a Gaussian function in real space, colored by its relative amplitude. Red indicates lower Gaussian amplitude and missing domains in that particle.

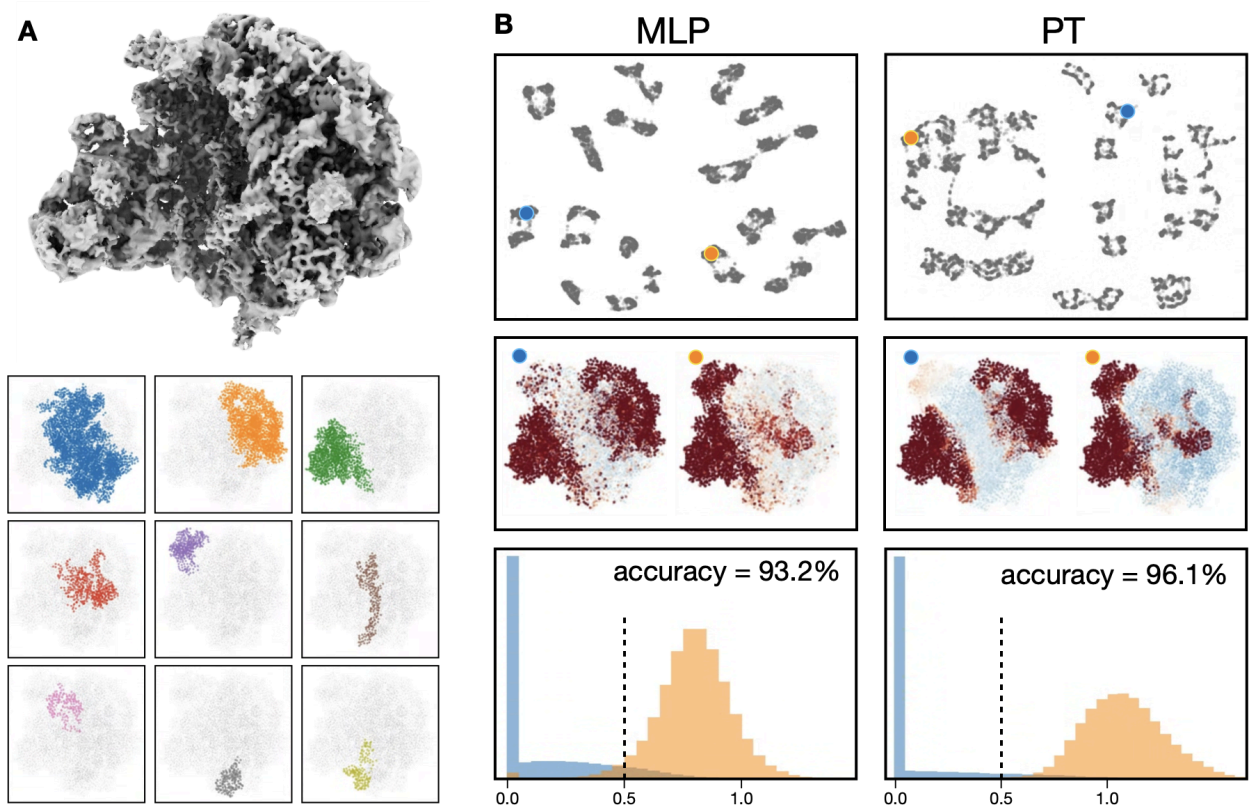

Figure 2. Comparison of MLP and PT-based decoder on simulated ribosome assembly dataset. (A) 3D structure of the 50S ribosome and the 9 segmented regions. Atoms of all regions, excluding the first one (blue), can be turned on/off when simulating each particle. (B) Comparison of MLP and PT-based implementation. Top: latent space of particles. Middle: decoder output of the corresponding points from the latent space, colored by relative amplitude of the Gaussian functions. Bottom: histogram of decoder amplitude output of Gaussian functions corresponding to atoms that were turned off in the particle (blue), and the atoms that were turned on (orange). The accuracy is reported with a cutoff threshold of 0.5.

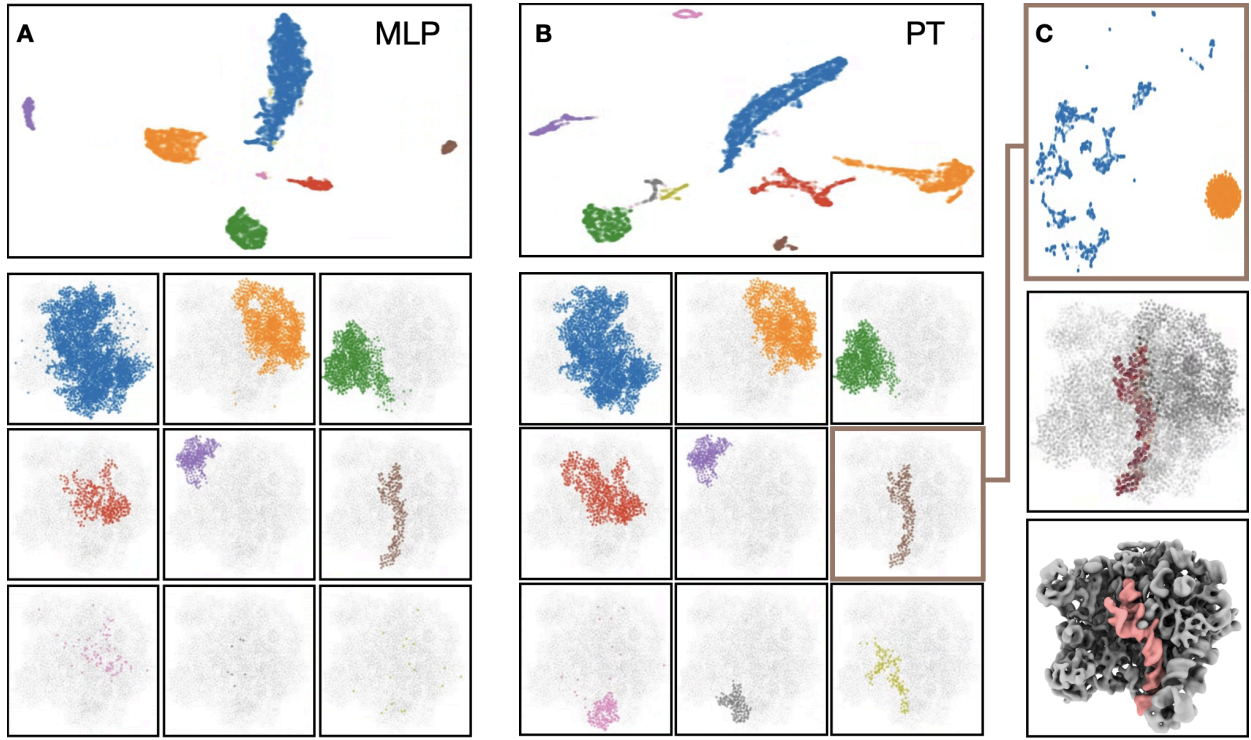

Figure 3. Segmenting the GMM using decoder output. (A-B) Comparison between MLP and PT-based implementation. Top: GMM embedded to a latent space according to decoder output. Each point represents one Gaussian function in the GMM. Points are colored using classes assigned by DB-Scan. Bottom: Gaussian functions corresponding to each class from the latent space. (C) Top: all particles embedded to a latent space using information of decoder output from one selected cluster in B, colored by K-means clustering result. Middle: difference of the decoder output from the two classes. Bottom: difference map of 3D reconstruction using particles of the two classes.

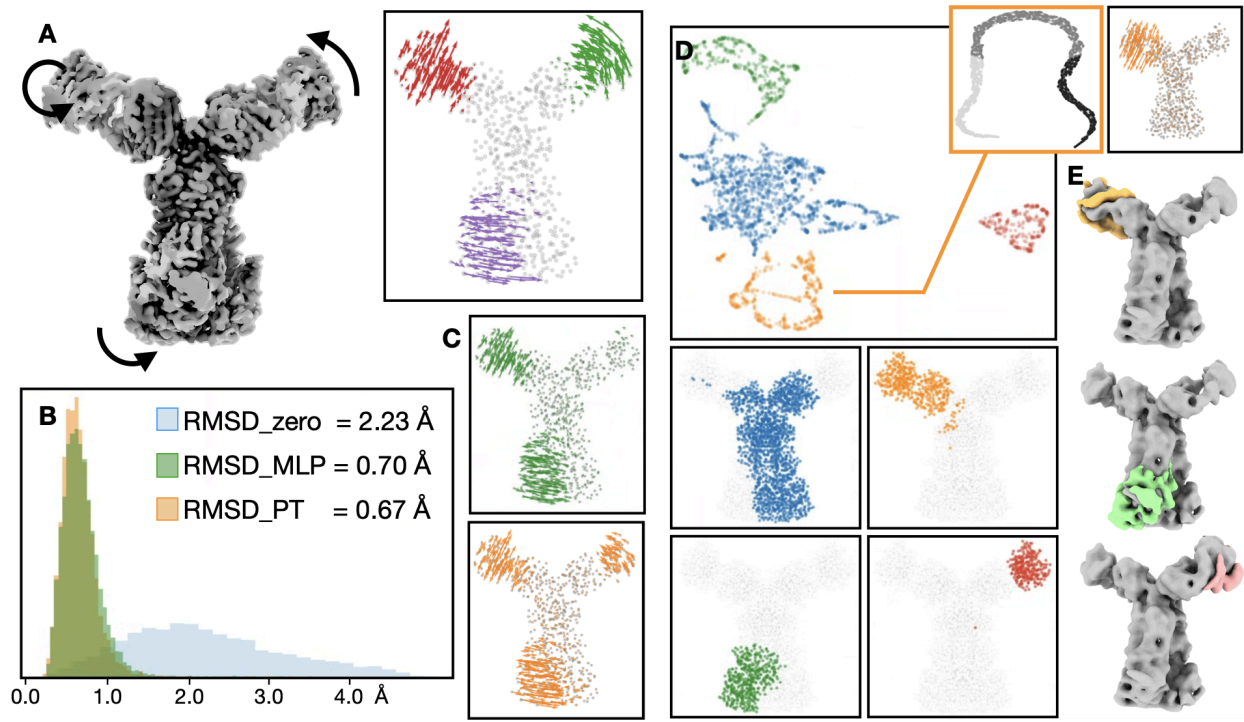

Figure 4. Comparison of MLP and PT-based decoder on continuous domain motion using a simulated dataset of an ABC-transporter. (A) Left: 3D structure of the transporter with Fab attachment, with arrows showing the simulated motion. Right: per-atom movement of one representative particle. (B) Histogram of RMSD between ground truth per-atom movement and: blue - rigid model with no motion; green - motion recovered by MLP model; orange - motion recovered by PT model. (C) Per-atom motion of the particle shown in A, recovered by: green - MLP; orange - PT-based model. (D) Top: latent space of GMMs generated from decoder output. Bottom: regions of the corresponding clusters. Right: latent space of particles using information of one selected region, and difference between decoder output from the first and last class. (E) Reconstruction of particles classified based on three latent spaces, each generated from information of one region from the GMM.

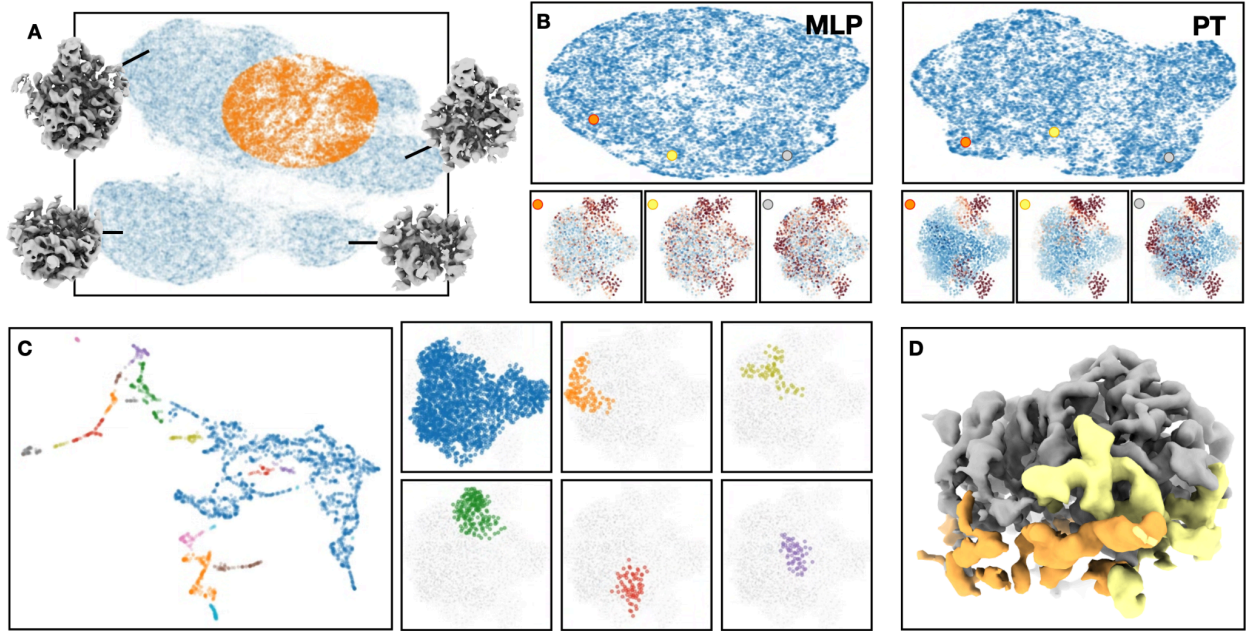

Figure 5. Application of PT-based model on real ribosome dataset. (A) Highlighted (orange) selected particles from the full dataset. The point cloud is generated by the existing MLP-based heterogeneity analysis of the full dataset, and the 3D reconstructions show some of the major classes from previous analyses. (B) Latent space of particles and decoder output using MLP and PT-based model on selected particles. (C) Left: latent space of the GMM, colored by clusters from DB-Scan. Right: regions of the six largest clusters. (D) 3D reconstructions of particle clusters from latent spaces focusing on two of the regions, showing the addition of RNA segments between the particle classes.